\documentclass[10pt,journal,compsoc]{IEEEtran}

\usepackage{subcaption}
\usepackage{graphicx}
\usepackage{grffile}
\usepackage{lipsum}
\usepackage[symbol]{footmisc}
\usepackage{amsmath,amssymb,amsfonts}
\usepackage{algorithmic}
\usepackage{textcomp}
\usepackage{xcolor}
\usepackage[linesnumbered,commentsnumbered,ruled,vlined]{algorithm2e}
\usepackage{caption}
\usepackage{xcolor, colortbl}
\usepackage{amsmath}
\usepackage{array}

\usepackage{booktabs} 

\usepackage{pifont}
\newcommand{\cmark}{\ding{51}}%
\newcommand{\xmark}{\ding{55}}%
\usepackage{lineno}
\usepackage{multirow}
\usepackage[usestackEOL]{stackengine}
\usepackage{rotating}
\usepackage[table]{xcolor}

\begin{document}
\newcommand{\best}[1]{\cellcolor{green!30} #1}
\newcommand{\worst}[1]{\cellcolor{red!30} #1}
\newcommand{\up}{\(\uparrow\)}
\newcommand{\down}{\(\downarrow\)}
\renewcommand{\thefootnote}{\fnsymbol{footnote}}

\newcommand{\tk}[1]{{\color{red}{(TK: #1)}}}
\newcommand{\red}[1]{{\color{red}{#1}}}


\title {How to Evaluate Distributed Coordination Systems? -- A Survey and Analysis}

\author{
\IEEEauthorblockN{Bekir Turkkan\IEEEauthorrefmark{1}\footnote{Contact }, Elvis Rodrigues\IEEEauthorrefmark{2}, Tevfik Kosar\IEEEauthorrefmark{2},\\ Aleksey Charapko\IEEEauthorrefmark{3}, Ailidani Ailijiang\IEEEauthorrefmark{4}, and Murat Demirbas\IEEEauthorrefmark{5}\\
\IEEEauthorblockA{\IEEEauthorrefmark{1}IBM Research, Yorktown Heights, NY}\\
\IEEEauthorblockA{\IEEEauthorrefmark{2}University at Buffalo, Buffalo, NY}\\
\IEEEauthorblockA{\IEEEauthorrefmark{3}University of New Hampshire, Durham, NH }\\
\IEEEauthorblockA{\IEEEauthorrefmark{4}Microsoft, Redmond, WA}\\
\IEEEauthorblockA{\IEEEauthorrefmark{5}MongoDB, New York City, NY}\\
 \IEEEcompsocitemizethanks{
        \IEEEcompsocthanksitem Contact authors: Bekir Turkkan (b.turkkan@ibm.com) and Tevfik Kosar (tkosar@buffalo.edu).}
}}

\IEEEtitleabstractindextext{%
\begin{abstract}
Coordination services and protocols are critical components of distributed systems and are essential for providing consistency, fault tolerance, and scalability. However, due to the lack of standard benchmarking and evaluation tools for distributed coordination services, coordination service developers/researchers either use a NoSQL standard benchmark and omit evaluating consistency, distribution, and fault tolerance; or create their own ad-hoc microbenchmarks and skip comparability with other services. In this study, we analyze and compare the evaluation mechanisms for known and widely used consensus algorithms, distributed coordination services, and distributed applications built on top of these services. We identify the most important requirements of distributed coordination service benchmarking, such as the metrics and parameters for the evaluation of the performance, scalability, availability, and consistency of these systems. Finally, we discuss why the existing benchmarks fail to address the complex requirements of distributed coordination system evaluation.

\end{abstract}

}

\maketitle
\thispagestyle{empty}

\section{Introduction}
\label{S:1}
Cloud and web-based big-data applications such as search engines, social networks, video streaming platforms, file-sharing tools, and the Internet of Things (IoT) are implemented as distributed systems, where a collection of nodes cooperate to achieve a common task for increased performance, availability, and scalability purposes. Developing, debugging, and evaluating such distributed systems have been a challenging task due to the coordination needed between the participating nodes of such systems. 

Distributed coordination is required for various purposes, including synchronization, locking, group membership, ownership, and reaching consensus. For the last three or four decades, the distributed systems community has developed protocols and services to achieve robust and scalable coordination in distributed applications. 
As one of the earliest efforts in this area, Paxos~\cite{Paxos} protocol was introduced by Lamport. The basic Paxos protocol gave rise to many variations and follow-up work, such as Disk Paxos~\cite{Disk-Paxos}, Cheap Paxos~\cite{Cheap-Paxos}, Fast Paxos~\cite{Fast-Paxos}, Generalized Paxos~\cite{Generalized-Paxos}, and Raft~\cite{RAFT}. To decrease the latency in Paxos communication and scale it to wide-area network (WAN) settings, many extensions were proposed, such as Mencius~\cite{Mencius}, Flexible Paxos (FPaxos)~\cite{Flexible-Paxos}, Egalitarian Paxos (E-Paxos)~\cite{E-paxos}, WPaxos~\cite{WPaxos}, and SwiftPaxos~\cite{ryabinin2024swiftpaxos}.

The difficulties in dealing with low-level Paxos and its variants in application development led to systems like ZooKeeper~\cite{Zookeeper}, Chubby~\cite{Chubby}, Tango~\cite{Tango}, and WanKeeper~\cite{Wankeeper}, which abstracted away distributed coordination and provided frequently used coordination primitives as a service to the application developers. Many distributed applications and especially high-demand distributed data stores such as Google's Spanner~\cite{Spanner}, Yahoo!'s PNUTS~\cite{PNUTS}, Apache's Mesos~\cite{Mesos} and Kafka~\cite{KAFKA}, and Twitter's Manhattan~\cite{Manhattan} were built on top of such coordination services, each requiring different levels of synchronization, consistency, and availability guarantees. 

Whether it is the development of a distributed coordination protocol or service or the development of an application employing these protocols/services, one of the significant challenges the developers face
is the lack of standard benchmarking tools that could provide a comprehensive evaluation of the coordination framework -- in terms of its performance, availability, scalability, and consistency guarantees. Currently, most developers use ad-hoc limited microbenchmarks to evaluate only a fraction of the functionality and use customized metrics and techniques. This results in a lack of comparability between coordination mechanisms and may lead to unfair claims of advantages over the competition.

In this paper, we study, analyze, and compare how different coordination systems are evaluated. More specifically, we try to answer the following questions: {\em What are the different evaluation metrics and parameters used by popular consensus algorithms, coordination services, and distributed applications? 
What are the major challenges and general requirements for achieving end-to-end benchmarking in distributed coordination systems? What functionality do the existing benchmarking systems provide, and where do they fall short in meeting the evaluation requirements of the distributed coordination systems?} 

The rest of this paper is organized as follows: Section 2 provides an analysis of the current evaluation practices for distributed coordination systems; Section 3 discusses the general requirements of a comprehensive benchmarking suite for distributed coordination; Section 4 presents the capabilities of existing benchmarking/testing frameworks in this area; Section 5 discusses the related work; and Section 6 concludes the paper.  

\section{Evaluation of Current Practices}
\label{S:3}

We have analyzed the existing distributed coordination systems under three main categories: (i) consensus algorithms; (ii) coordination services; and (iii) distributed applications. The relevance, recency, and impact of the systems were considered during the selection process.

{\bf (i) Consensus Algorithms.} These are the low-level algorithms and protocols designed to provide the necessary primitives for distributed coordination. Most of the systems in this category are variations of the original Paxos protocol, with improvements either in performance, availability, or scalability. The systems we have studied under this category include Mencius~\cite{Mencius}, FPaxos~\cite{Flexible-Paxos}, Raft~\cite{RAFT}, Multi-Paxos~\cite{Multi-Paxos}, Hybrid-Paxos~\cite{Hybrid-Paxos}, Egalitarian Paxos (E-Paxos)~\cite{E-paxos}, $M^2$ Paxos~\cite{M2-Paxos}, Bizur~\cite{Bizur}, ZAB~\cite{ZAB}, WPaxos~\cite{WPaxos}, SwiftPaxos~\cite{ryabinin2024swiftpaxos}, Omni-Paxos~\cite{omnipaxos}, and Hydra~\cite{Hydra}. 

{\bf (ii) Coordination Services.} These are high-level services that form a level of abstraction to hide the difficulties in dealing with low-level consensus algorithms and provide the frequently used coordination primitives as a service to the application developers. The systems studied under this category include ZooKeeper~\cite{Zookeeper}, Tango~\cite{Tango}, Calvin~\cite{Calvin}, WanKeeper~\cite{Wankeeper}, ZooNet~\cite{Zoonet}, Boki~\cite{Boki}, FlexLog~\cite{Flexlog}, SplitFT~\cite{SplitFT}, Fabric~\cite{HyperledgerFabric}, and Narwhal~\cite{Narwhal}.

{\bf (iii) Distributed Applications.} The final category consists of distributed applications built on such consensus algorithms or coordination services, each requiring different levels of synchronization, consistency, and availability guarantees. The systems we have studied under this category are: Spanner~\cite{Spanner}, DistributedLog~\cite{DistributedLog}, PNUTS~\cite{PNUTS}, COPS~\cite{COPS}, Cockroach DB \cite{cockroach}, OceanBase~\cite{oceanbase}, and ScalarDB~\cite{ScalarDB}.

\subsection{Topologies and Experimental Setups}
\label{Setups}
\begin{figure*}[t]
\centering
\includegraphics[width=0.7\textwidth]{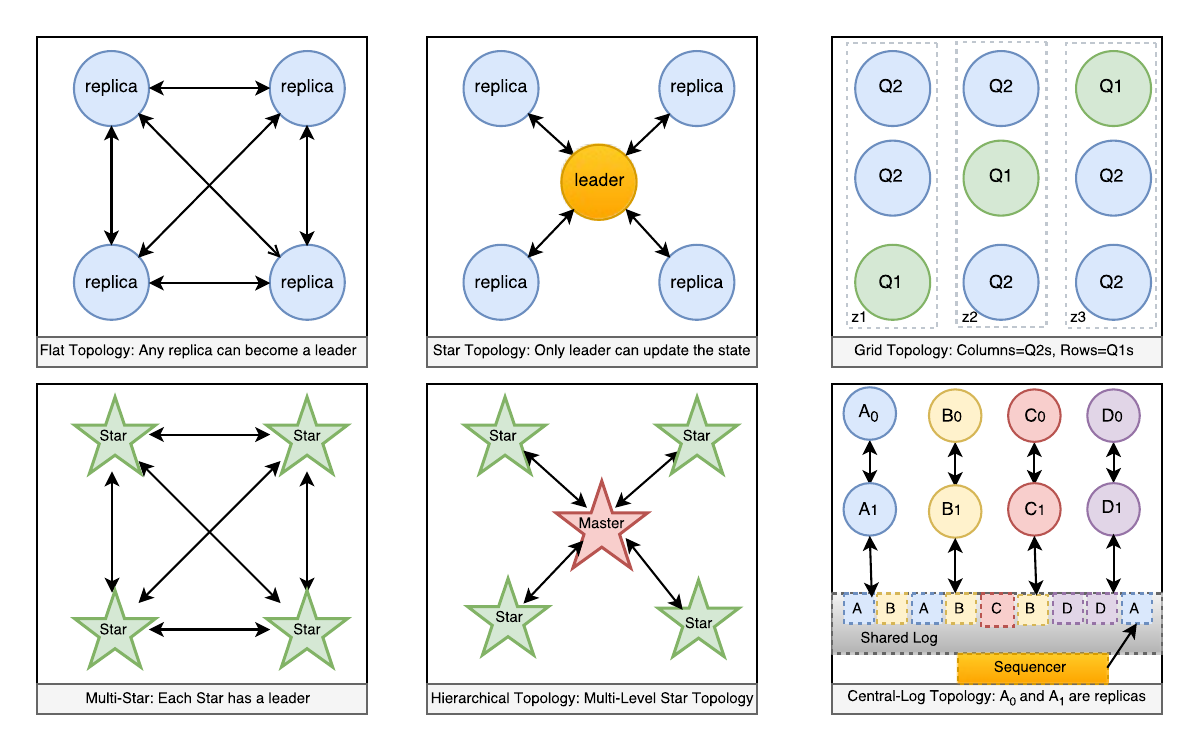}
\vspace{-2mm}
\caption{Different topologies used in system evaluations.}
\label{fig:topologies}
\end{figure*}

In the evaluation of distributed coordination systems, different experimental setups and topologies are used.  
We group these topologies under six main categories: (i) flat topology; (ii) star topology; (iii) multi-star topology; (iv) hierarchical topology; (v) grid topology; and (vi) central-log topology. To categorize the experimental setups in these topologies, the implementation and design principles of the studied systems have been analyzed. The way to create quorums and the way to process requests are considered as the two main factors.
Figure \ref{fig:topologies} illustrates these topologies.

{\em Flat topology} is used in multi-leader or leaderless systems that allow concurrent updates like Mencius and E-Paxos. Although their way of assigning a replica to be the leader of a request differs, the topology represents the architecture of both systems.
Similarly, the {\em star topology} is used in single leader protocols, whether the system relies on strong leadership like ZooKeeper or changes the leader based on the outstanding requests at replicas, as in Hybrid-Paxos. 
{\em Multi-Star topology} is used in systems with multiple quorums, each in a star topology, while leaders of every quorum communicate with each other and form a flat topology together (no hierarchy). Systems with partitioned data and access locality may use Multi-Star topology for better read performance or improved fault tolerance as in ZooNet and $M^2$ Paxos. Systems with dynamic quorums, according to their data partition and replication requirements, as in Spanner, may also use this topology. 
{\em Hierarchical topology} can be defined as a multi-star topology with a hierarchy between the leaders. It is used in WanKeeper mainly to control the distribution of ownership of data objects in wide-area networks, as the master is the only replica to change the ownership. 
FPaxos uses {\em grid topology} and grid quorum to improve the performance in a cluster, while WPaxos benefits from the same to decrease the wide-area networking (WAN) communication overhead for the write operations.
{\em Central-log topology} is used by systems like Tango, Boki, and Calvin, which keep the execution order of transactions in a durable, consistent, and fault-tolerant shared log.

\begin{table*}
\centering
\scriptsize
\strutlongstacks{T} 
\begin{tabular}{l|l| l| l| l| l|l|l|}
\cline{2-8}
& \multirow{2}{*} {\textbf{System}}  & \textbf{\Longunderstack{\# of \\Regions}}&\textbf{\Longunderstack{\# of \\Servers}} & \textbf{\Longunderstack{\# of\\ Clients}}  & \multirow{2}{*}{\textbf{Topology}} &\multirow{2}{*}{\textbf{Testbed}} &\multirow{2}{*}{\textbf{Benchmark}}\\
\hline \hline
\multirow{13}{*}{\textbf{\begin{turn}{90}Algorithms\end{turn}}}
& Mencius~\cite{Mencius} &3,5,7 &3,5,7 &3,5,7  &Flat & DETER &Microbenchmark\\
& FPaxos ~\cite{Flexible-Paxos}  &1  &5,8  & ---  &Grid & Custom &Microbenchmark\\
& Raft ~\cite{RAFT}  &1  &5,9  & ---  &Star & Custom &Microbenchmark\\
& Multi-Paxos ~\cite{Multi-Paxos} &1  &5 &1,4,10,20  &Star & Custom &Microbenchmark\\
&Hybrid Paxos ~\cite{Hybrid-Paxos} &5,7,11,21 &5,7,11,21 &10,20,100,1000  &Star &Emulab &Microbenchmark\\
&E-Paxos~\cite{E-paxos} &1,3,5 &3,5 &50  &Flat &EC2 &Microbenchmark\\
&$M^2$ Paxos ~\cite{M2-Paxos}&1 &5,11,49 &64/server  &Multi-Star &EC2 &Microbenchmark\\ 
&Bizur ~\cite{Bizur}&1 &3 & 1  &Flat & Custom &Microbenchmark\\
&ZAB ~\cite{ZAB} &1  &3-13 & ---  &Star & Custom &Microbenchmark\\
&WPaxos ~\cite{WPaxos}&5 &15 &5   &Grid & EC2 &Microbenchmark\\
&SwiftPaxos ~\cite{ryabinin2024swiftpaxos} &13 &5 &1000,5000 &Star &EC2 &YCSB ~\cite{YCSB} \\
&Omni-Paxos ~\cite{omnipaxos} &1,3&3,5&1&Flat&Google C.E.&Microbenchmark\\
&Hydra ~\cite{Hydra} &1 &8 &--- &\Longunderstack[l]{Star\\Hierarchical} &Custom &\Longunderstack[l]{Microbenchmark\\ YCSB+T \cite{ycsb+t}} \\
 \hline
\multirow{12}{*}{\textbf{\begin{turn}{90}\Longunderstack{Coordination\\Services}\end{turn}}}
&ZooKeeper ~\cite{Zookeeper}&1  &3-13 &250  &Star &Custom &Microbenchmark\\
&Tango ~\cite{Tango}&1 &18 &18  &Central-Log &Custom &YCSB ~\cite{YCSB}\\
&Calvin ~\cite{Calvin} & 1 &4,8,0-100 & ---   &Central-Log &EC2 &TPC-C ~\cite{TPC-C}\\
&WanKeeper ~\cite{Wankeeper} &3 &3 &1,2  &Hierarchical &EC2 &YCSB ~\cite{YCSB}\\
&ZooNet ~\cite{Zoonet}  &2 &8 &60 &Multi-Star &Google C.E. &Microbenchmark\\ 
&Boki ~\cite{Boki}&1 &11, 64 &64, 96, 128, 192 &Central-Log &EC2 &\Longunderstack[l]{Microbenchmark \\ DeathStar \cite{DeathStar}} \\
&FlexLog ~\cite{Flexlog}&1 &1-6 &1 &Central-Log &Custom &Microbenchmark\\
&SplitFT ~\cite{SplitFT} &1 &1, 20 &5, 43 &Star &CloudLab &\Longunderstack[l]{Microbenchmark\\ YCSB \cite{YCSB}} \\
&Fabric ~\cite{HyperledgerFabric} & 1, 2, 5 & 15-110 & --- & Flat & IBM Cloud & Microbenchmark \\
&Narwhal ~\cite{Narwhal} &5 &8---51 &1---10 &Flat &EC2 &Microbenchmark \\
 \hline
\multirow{10}{*}{\textbf{\begin{turn}{90}Applications\end{turn}}}
&Spanner ~\cite{Spanner} &1 &1,3,5 & 100   &Multi-Star &Custom &Microbenchmark\\
&DistibutedLog ~\cite{DistributedLog}&2 &20  &---  &Star &Custom &Microbenchmark\\
&PNUTS ~\cite{PNUTS} &3 &3 &300   &Multi-Star & Custom &Microbenchmark\\
&COPS ~\cite{COPS} &1 &2,4,8,16,32 &2,1024/server    &Multi-Star &VICCI &Microbenchmark\\
&CockroachDB~\cite{cockroach} &1&3,30&---&Flat &Custom &\Longunderstack{TPC-C \cite{TPC-C} \\ Jepsen \cite{jepsen}}\\
&OceanBase ~\cite{oceanbase} &3 &1557 &360000/server &Flat&Alibaba ECS&TPC-C \cite{TPC-C}\\ 
&ScalarDB ~\cite{ScalarDB} &1 &2, 203-1015 &4, 8 &Star &EC2 &\Longunderstack[l]{Elle \cite{Elle}\\ YCSB \cite{YCSB}\\ TPC-C \cite{TPC-C}}\\
\hline
\end{tabular}
\caption{Evaluation setup of the studied systems ( "---" means details are not specified).}
\label{tab:eval_setup}
\end{table*}


In Table~\ref{tab:eval_setup}, the experimental setups of the studied systems are given. The number of regions field shows the level of geo-distribution. If systems are deployed in real-world computing environments such as public clouds (as in ZooNet and WanKeeper) or geographically distributed private networks (as in PNUTS), and the round-trip time (RTT) between distributed replicas is greater than 20ms, it is included as a region. Similarly, for evaluations in controlled environments, if RTT between modeled regions is larger than 20ms, as in Mencius and Hybrid-Paxos, it is also accepted as a separate region. However, if all the servers are located in the same placement group (as in $M^2$ Paxos) or clusters in close proximity (as in Spanner and Calvin), it is considered a single region deployment even if the system design parameters support wide-area network deployments. The number of server fields represents the level of replication. The source of the produced workload is defined as the client, and the total number of workload sources is given as the number of clients.  

Systems apply different ways of creating client tasks. The general approach is to utilize a separate machine to create the required workload. For single-cluster evaluations using a separate machine, one or multiple threads are used as clients, as in Bizur and Multi-Paxos, respectively. In WAN deployments, clients are distributed to the regions uniformly, as in Mencius, E-Paxos, and WPaxos. On the other hand, ZAB, FPaxos, and DistributedLog do not implement external clients. Similarly, $M^2$ Paxos uses separate threads on the replica servers to generate the workload. ZooKeeper (35 servers for 250 clients) and ZooNet (2 servers for 60 clients) combine these approaches and use separate machines with multiple client threads. We provide testbed information to clarify whether the evaluations were simulated under a controlled environment like Emulab \cite{Emulab} and DETER \cite{DETER} or performed on real computing environments like CloudLab \cite{CloudLab}, Amazon EC2 \cite{EC2}, Alibaba ECS \cite{alibabaECS}, and Google Compute Engines \cite{GCE}. This information is important to understand the involvement of external factors in evaluations, such as variance in routing overhead, as it is fully deterministic in controlled environments, while you have limited control in real computing environments. 

Custom testbeds can be considered in between as they are closer to controlled environments for single-region deployments and closer to cloud computing environments for geo-distributed deployments. Under the benchmark field, if a standard benchmark is not used for the evaluations, it is listed as "Microbenchmark." Different implementations in evaluations of these custom microbenchmarks are given under the related metrics. Table~\ref {tab:eval_setup} lists all of the experimental setups used in the evaluations of studied systems together, although some of them change their experimental setup for different metrics. For instance, Mencius uses 3 servers in 3 regions for performance evaluations; however, it uses 3,5,7 servers in 3,5,7 regions for the scalability benchmarks. Similarly, Calvin studies conflicting command performance with 4 and 8 replicas while using 100 replicas to conduct scalability experiments. If different experimental setups of the same system are important, they are explained in detail in the section of the corresponding metric. For any of the fields in the table, if the details are not given explicitly in the related published materials, they are marked as "Not Specified" in the table.

\vspace{-2mm}
\subsection{Metrics}
\label{S:Metrics}
The majority of the protocols and systems we have studied are evaluated for several different metrics. These metrics describe various aspects of system behavior under the workload, such as performance, scalability, availability, and consistency. Despite the lack of unity in the benchmarking of the systems, many authors have chosen to use the same metrics for their evaluations. In the rest of this section, we will describe the benchmarking metrics used in the literature and how these metrics allow researchers to evaluate their systems.
Table~\ref{tab:metrics} summarizes our findings.

\begin{table*}[h]
\centering
\begin{tabular}{l|l|l|l|l|l|l|l|l|}
\cline{2-9}
& \multirow{2}{*}{\textbf{System}}&  \multicolumn{2}{|c|}{\textbf{Performance}} & \multicolumn{2}{|c|}{\textbf{ Scalability}} & \multicolumn{2}{|c|}{\textbf{Availability}} & \multirow{2}{*}{\textbf{Consistency}}\\
\cline{3-8}
 & & \textbf{Throughput} & \textbf{Latency} & \textbf{Server} & \textbf{Client} & \textbf{Failure}  & \textbf{Partition} & \\ \hline \hline
\multirow{13}{*}{\textbf{\begin{turn}{90}Algorithms\end{turn}}} 
&Mencius~\cite{Mencius} &\cmark &\cmark &\cmark & &\cmark &\cmark &\\
&FPaxos ~\cite{Flexible-Paxos}&\cmark &\cmark & & & & & \\
&Raft ~\cite{RAFT}& &\cmark & & &\cmark & & \\
&Multi-Paxos~\cite{Multi-Paxos} & \cmark & & &\cmark & & &\cmark \\
&Hybrid Paxos~\cite{Hybrid-Paxos} &\cmark &\cmark &\cmark &\cmark & & & \\
&E-Paxos~\cite{E-paxos} & \cmark & \cmark & \cmark & &\cmark &\cmark &\\
&$M^2$ Paxos~\cite{M2-Paxos} &\cmark &\cmark &\cmark & & & & \\
&Bizur~\cite{Bizur}  &\cmark &\cmark & & &\cmark & &\cmark \\
&ZAB ~\cite{ZAB} &\cmark &\cmark &\cmark & & & & \\
&WPaxos ~\cite{WPaxos} &\cmark &\cmark &\cmark & &\cmark & & \\ 
&SwiftPaxos ~\cite{ryabinin2024swiftpaxos} &\cmark &\cmark & &\cmark & & & \\
&Omni-Paxos ~\cite{omnipaxos} &\cmark & &\cmark &\cmark &\cmark &\cmark & \\
&Hydra ~\cite{Hydra} &\cmark &\cmark &\cmark & &\cmark & & \\
\hline
\multirow{10}{*}{\textbf{\begin{turn}{90}\Longunderstack{Coordination\\Services}\end{turn}}}
&ZooKeeper ~\cite{Zookeeper} &\cmark &\cmark &\cmark &\cmark &\cmark & & \\
&Tango ~\cite{Tango}&\cmark &\cmark &\cmark &\cmark & & &\\
&Calvin ~\cite{Calvin} &\cmark & &\cmark & & & &\\
&WanKeeper ~\cite{Wankeeper} &\cmark &\cmark &\cmark & & & & \\
&ZooNet~\cite{Zoonet}  &\cmark & &\cmark &\cmark & & &\cmark \\
&Boki~\cite{Boki} &\cmark &\cmark &\cmark & & & & \\
&FlexLog~\cite{Flexlog} &\cmark &\cmark &\cmark & &\cmark & & \\
&SplitFT~\cite{SplitFT} &\cmark &\cmark & & &\cmark & & \\
&Fabric~\cite{HyperledgerFabric} &\cmark &\cmark &\cmark & & & & \\
&Narwhal~\cite{Narwhal} &\cmark &\cmark &\cmark &\cmark &\cmark & & \\
 \hline
\multirow{7}{*}{\textbf{\begin{turn}{90}Applications\end{turn}}}
&Spanner ~\cite{Spanner} &\cmark &\cmark &\cmark &\cmark &\cmark &\cmark & \\
&DistributedLog ~\cite{DistributedLog} &\cmark &\cmark &\cmark &\cmark & & &\\
&PNUTS ~\cite{PNUTS} & &\cmark & &\cmark & & & \\
&COPS ~\cite{COPS} &\cmark &\cmark &\cmark &\cmark & & &  \\
&CockroachDB ~\cite{cockroach} &\cmark &\cmark &\cmark & & & &\cmark \\
&OceanBase ~\cite{oceanbase} &\cmark&\cmark&\cmark&\cmark& & & \\
&ScalarDB~\cite{ScalarDB} &\cmark &\cmark &\cmark &\cmark & & &\cmark \\
\hline
\end{tabular}
\caption{Systems evaluations based on metrics (\cmark : measured). }
\label{tab:metrics}
\end{table*}

\subsubsection{\textbf{Performance}}
\label{S:Performance}
\hfill\\
Performance evaluation is the most common type of system benchmarking. Studying the performance typically involves pushing some significant workload through the system and measuring the latency and the throughput the system was able to sustain. 

\begin{table*}[h]
\centering
\begin{tabular}{l|l| l|l| l| l| l|}
\cline{2-7}
& \multirow{2}{*}{\textbf{System}} 
&\textbf{\Longunderstack{Write \\Ratio \%}}
& \textbf{\Longunderstack{Data \\Access\\Overlap \%}} & \textbf{\Longunderstack{Access \\Locality \%}}& \textbf{\Longunderstack{\# of\\ Objects}}& \textbf{\Longunderstack{Size of\\ Objects}}\\
\hline \hline
\multirow{13}{*}{\textbf{\begin{turn}{90}Algorithms\end{turn}}}
& Mencius~\cite{Mencius} & 50 &100 &NS  &16,128,1024 &6B,4KB \\
& FPaxos ~\cite{Flexible-Paxos} & NS &\xmark &\xmark &NS &64B\\
& Raft ~\cite{RAFT} & \xmark &\xmark &\xmark &\xmark &\xmark\\
& Multi-Paxos ~\cite{Multi-Paxos} &100 &100 &100 &NS &5B,8KB, 32KB\\
&Hybrid Paxos ~\cite{Hybrid-Paxos} &100 &100 &\xmark &1 &NS \\
&E-Paxos~\cite{E-paxos} &100 &100 &NS &NS &16B,1KB \\
&$M^2$ Paxos ~\cite{M2-Paxos} &NS &2-100 &0,100 &1,100,1000 &16B \\ 
&Bizur ~\cite{Bizur} &100 &NS &\xmark &1-$2^{13}$ &50B \\
&ZAB ~\cite{ZAB} &NS &100 &\xmark &NS &1KB \\
&WPaxos ~\cite{WPaxos} &100 &NS &70, 90 &1000 &NS \\
&SwiftPaxos ~\cite{ryabinin2024swiftpaxos} &0,5,20 &0-100 &\xmark &$10^6$ &1KB-8KB \\
&Omni-Paxos ~\cite{omnipaxos} &100 &100 &\xmark &500, 5K, 50K &8B \\
&Hydra ~\cite{Hydra}&100, 50 &0-100 &NS &NS &NS \\
 \hline
\multirow{11}{*}{\textbf{\begin{turn}{90}\Longunderstack{Coordination \\Services}\end{turn}}}
&ZooKeeper ~\cite{Zookeeper} &0-100 &100 &\xmark &NS &1KB \\
&Tango ~\cite{Tango} &\Longunderstack[l]{0,10,50\\90,100} &100 &0-100 &10-10M &4KB \\
&Calvin ~\cite{Calvin} &0-34~\cite{TPCC_char} &NS &90 &NS &NS  \\
&WanKeeper ~\cite{Wankeeper} &0-100 &0-100 &0,100 &NS &NS \\
&ZooNet ~\cite{Zoonet} &\Longunderstack{75,50,10,1,0} &NS  &NS &NS &1KB \\ 
&Boki ~\cite{Boki} &0-100 &0-100 &NS &NS &1KB \\
&FlexLog ~\cite{Flexlog} &5-95 &0-100 &NS &NS &1KB, 64B - 8KB \\
&SplitFT ~\cite{SplitFT} &0-100 &NS &NS & $10^7$-$10^8$ &128B - 8KB \\
&Fabric ~\cite{HyperledgerFabric} &50-100 &100 &NS & $2^{19}$ &0.5MB-4MB \\
&Narwhal ~\cite{Narwhal} &100 &NS &NS & NS &0.5MB \\
 \hline
\multirow{7}{*}{\textbf{\begin{turn}{90}Applications\end{turn}}}
&Spanner ~\cite{Spanner} &0,100 &NS &NS &NS &4KB \\
&DistibutedLog ~\cite{DistributedLog} &NS &100 &NS &NS &1KB \\
&PNUTS ~\cite{PNUTS} &10,0-50 &NS &80 &NS &NS \\
&COPS ~\cite{COPS} &50,25 &100 &0-100 &$2^{18}$, 512/client &1B \\
&CockroachDB ~\cite{cockroach} &0-34~\cite{TPCC_char} &100 &90 &1000, 10000 &NS \\
&OceanBase ~\cite{oceanbase} &0-34~\cite{TPCC_char} &NS &NS & $2^{40}$ &4KB - 8KB\\
&ScalarDB ~\cite{ScalarDB} &0-50~\cite{YCSB, TPCC_char} &0-100 &NS &$2^{27}$ &1KB \\
\hline
\end{tabular}
\caption{Workload parameters used in performance evaluations (\xmark: Not available or applicable for experiments conducted on this system; NS: Details are not specified).
}
\label{tab:performance}
\end{table*}

{\bf Throughput} measures how many requests or commands a system can process in a unit of time. Higher throughput capabilities allow the system to process more requests and larger amounts of data. To measure the maximum throughput, systems are saturated with an increasing workload until they reach the server's CPU or network capacity.

{\bf Latency} is a measure of request execution time. It is often measured over a significant number of requests, allowing the calculation of average, median, and various percentile's for the metric. Low latency is desirable since it means the system spends little processing power or I/O capacity to handle a request. Latency for a given system is often correlated with throughput, as low request latency allows us to push more requests through the system in unit time.

Latency distribution across the requests is also very important.  For instance, the average latency measurement can give some sense of the performance; however, it does not describe the full picture that includes all requests the system serves. It may be the case that an average or median latency is low while some significant portion of requests performs poorly. Most systems give the results only for average, median, or aggregated latency without providing any information about the tail latency. This makes it more difficult to understand how a system works in performance corner cases. To this matter, providing the distribution of the cumulative latencies as in WPaxos and OceanBase or giving median or average and high percentile latency together as in E-Paxos, Bizur, WPaxos, and Boki yields a more comprehensive understanding of the system’s performance.

Systems follow different approaches for evaluating performance. Although the majority of the systems evaluate the throughput and latency together, some systems, such as Calvin, ZooNet, Multi-Paxos, and Omni-Paxos, measure only throughput, and systems such as ZooKeeper and Spanner evaluate the throughput and latency separately. Evaluating latency and throughput together is essential to understanding the system's overall performance. While the throughput is mainly related to the system's processing capacity, latency is more affected by I/O operations \cite{latency-throughput}.

The performance of any distributed coordination system highly depends on workload characteristics such as read/write ratio, data access overlap, access locality, the number of data objects (size of data pool), and the size of data objects. In table \ref{tab:performance}, we list how the systems tuned their workloads for these parameters.

\paragraph{Read/Write Ratio}
\label{S:RWratio}
Read/Write Ratio is a fundamental parameter to show the system's performance under different use case scenarios. Since many systems handle write and read requests differently, and the percentage of write and read requests is application-specific, it plays a crucial role in showing the applicability of the systems for various purposes. For instance, systems designed to be used for different aspects of distributed coordination, like ZooKeeper or WanKeeper, show the performance evaluation continuously while varying the write ratio from 0 to 100\%. Similarly, Spanner measures the read and write performance separately since read requests can be handled by replicas, but writes need to be done by the leader. Besides being handled by the leader, writes require replication, which increases commit latency. To measure the true performance of replication, Multi-Paxos also uses 100\% write operations. Due to the differences in processing read requests between systems used for comparison in the evaluations, Bizur uses only write operations. E-Paxos and Hybrid Paxos also use 100\% writes since they focus on handling conflicting commands, and read requests do not produce conflicts. Some works like Calvin, CockroachDB, and OceanBase do not explicitly provide a write ratio for their evaluations but use standard benchmarks like TPC-C \cite{TPCC_char} and YCSB \cite{YCSB} with known write ratios.

\paragraph{Data Access Overlap}
\label{S:DataAccessOverlap}
Data access overlap may have a major impact on the throughput and latency in multi-leader systems and protocols. It can be defined as the percentage of the key space shared by all clients. In this matter, data access overlap is 100\% if the entire key space is shared and all clients are allowed to access any key. On the other hand, 
data access overlap is 0\% if the key space is partitioned for each client and clients are accessing only the keys in their partition. For single leader algorithms providing strong consistency like Multi-Paxos, ZAB, Chubby, and ZooKeeper, varying data access overlap has no impact on the overall performance since all write requests are treated in the same way. 
Data access overlap is an important factor for the performance of multi-leader algorithms/systems that allow concurrent update operations. Receiving different execution orders at different replicas is considered a command conflict. A command conflict is usually resolved by serializing the requests from different clients to have the same execution order or by checking the dependency between the requests to make sure the order of execution does not lead to different final states of the system. Mencius and E-Paxos can be considered under this category and they evaluate their system under the worst condition with 100\% data access overlap. Hybrid-Paxos is also considered to be evaluated with 100\% data access overlap since all clients are updating the same data object (one bank account). For these systems, data access overlap can also result in command collisions, as described in \ref{S:CommandCollisionRatio}, which occurs when multiple servers update the same data object concurrently. Systems that use ownership to control access to data objects are also affected by the conflict ratio due to ownership migration or remote requests. Systems using static ownership, like ZooNet, forward all requests to the remote server, which is possibly located in a geographically far region, and they are affected by network delays for requests of a non-owned data object. The use of buckets in Bizur is similar. Systems with dynamic ownership are either affected by exhibiting longer latency for remote requests or performing costly ownership migration, as in $M^2$ Paxos, WPaxos, WanKeeper, and PNUTS. 

\paragraph{Command Collision Ratio}
\label{S:CommandCollisionRatio}
`Command collision occurs in multi-leader systems when more than one leader is trying to operate on the same data object at the same time. For example, Mencius implements a simple replicated register service and evaluates the effect of command collision by changing the number of registers available for the clients. 
The lower number of registers leads to a greater collision rate. Similarly, Hybrid Paxos and E-Paxos evaluate the collisions by changing the rate of commands updating the same key. Hybrid Paxos measures the latency by changing the withdraw operation ratio from 0\% to 100\%, where 100\% represents the case of a 100\% command collision rate. E-Paxos evaluates command collision for 0\% and 2\% as likely cases and 25\% and 100\% as extreme cases. 

\paragraph{Access Locality}
\label{S:AccessLocality}
 Access locality is defined as the likelihood of clients accessing a specific part of the key space. If there is a 70\% access locality, it represents the case that 70\% of the requests are related to the same part of the key space, possibly the same region, and the other 30\% of the requests are related to the rest of the key space. Access locality is more of a matter of the distribution of client requests rather than the sharing of the key space among clients, which is defined as data access overlap. Clients may have different access localities while preserving 100\% data access overlap as in the above example. 
 This parameter has a big impact on performance in systems that use object ownership to parallelize request execution. 

Access locality may not necessarily correspond to geographical partitioning of data, and in more general terms, it can be seen as the probability of accessing some preferred subset of keys, as illustrated in $M^2$ Paxos. WPaxos and COPS adjust the access locality by distributing the data objects uniformly in data groups and changing the client access rate for the data groups. While COPS changes the access locality for each client from 100\% to 0\%, WPaxos evaluates the access locality for 70\% and 90\%. $M^2$ Paxos measures the performance for the worst and best conditions as 0\% and 100\% access locality. 
ZooNet and PNUTS distribute the ownership of the data objects to adjust the level of local execution so their access locality rates correspond to the rate of local execution. Similarly, Calvin partitions its data to multiple datacenters and multiple machines in the same datacenter. Calvin's evaluations exhibit a 90\% access locality since it accesses the data objects stored on the same machine for 90\% of the cases. PNUTS also examines the effect of non-uniform access on data objects by using Zipfian distribution~\cite{zipfian} with varying Zipf factors, which results in different access locality rates.   

\paragraph{Number of Data Objects} 
\label{S:NumberOfDataObjects}
The number of data objects is usually used to tune command conflicts and command collisions. The higher number of data objects usually results in fewer conflicts and fewer collisions. Only Mencius, WPaxos, Bizur, COPS, $M^2$ Paxos, Omni-Paxos, ScalarDB, and OceanBase specify the number of data objects that could be used for adjusting the ratio of conflicting and possibly colliding commands.

\paragraph{Size of Objects} 
\label{S:SizeOfObjects}
The size of objects, as another parameter of performance evaluation, is usually used as a factor of the saturation type. To make the workload CPU-bound, the size is kept as small as possible, and for the network-bound workloads, system evaluation may use larger objects. It is also related to the data models and the expected use of the systems. ZooKeeper, Tango, Spanner, Boki, FlexLog, and ZooNet use 1KB or 4KB-sized data objects, as these are more suitable for their data model. Mencius and Multi-Paxos evaluate the effect of the data object size to evaluate the system for both CPU and network-bound cases.

\subsubsection{\textbf{Scalability}}
\label{S:Scalability}
\hfill\\
The ability of any system to perform well in a wide range of workload parameters and configurations is essential. The scalability measured by the system is of two distinct types: {\bf workload scalability} and {\bf system scalability}. Both types are often evaluated by measuring the changes in performance as some parameters are controlled. The parameters important for scalability evaluation are often those that describe the experimental testbed and can be seen in Table \ref{tab:eval_setup}.

{\bf Workload scalability} is often evaluated by fixing the deployment size and topology while increasing the amount of work the system needs to perform and measuring the performance at each workload intensity. Systems control the throughput as a measure of workload intensity. At low-intensity workloads, systems normally show stable performance that does not degrade drastically with small increases in the number of requests. However, as the number of requests sent to the system increases, it eventually reaches a saturation point where even a small increase in the number of requests results in drastic degradation in latency and the system's ability to process higher throughput. Systems using small data objects usually reach the saturation point due to CPU limitations, and systems using larger file sizes are more affected by network boundaries.

{\bf System scalability} is often measured by keeping workload parameters constant and changing the size of deployment by either changing the number of servers/replicas or by changing the number of regions for wide area network systems. System scalability is important for large data-driven systems that must scale the processing and storage capacity well.

\paragraph{Number of Clients}
\label{S:NumberOfClients}
To evaluate workload scalability, a certain amount of workload should be created until systems reach their saturation point. The amount of workload can easily be adjusted by either changing the number of clients or the number of requests per client. Hybrid-Paxos, E-Paxos, Tango, ScalarDB, and PNUTS increase the number of concurrent clients with a certain amount of workload produced by each client for the workload scalability evaluations. $M^2$ Paxos also increases the number of clients, but it is more related to producing a new workload for the newly added server. Thus, it is not considered directly related to workload scalability evaluation. In some cases, clients may be throttled down, and changing the degree of throttling can be used to control the workload intensity, as it is done in WPaxos. 

\paragraph{Number of Servers/Replicas}
\label{S:NumberOfServers}
Depending on the underlying algorithms, systems differ in how they scale with the increasing number of servers/replicas. Paxos derivations generally provide low performance for a higher number of replicas. Some systems, such as ZooKeeper, however, can provide increased read throughput as the number of replicas grows since these systems do not put a single leader in a read path and allow reading directly from replicas. Some works like Eris~\cite{Eris}, NOPaxos~\cite{NOPaxos}, and HydraPaxos~\cite{Hydra} sidestep consensus with in-network sequencing and groupcasts to achieve better performance with more replicas than consensus-based derivations. Increasing the number of servers/replicas does not always increase the aggregated throughput linearly due to the communication and replication overheads for most systems. 
To evaluate system scalability as the number of servers/replicas increases, evaluations often resort to CPU-bound workloads, as was done by $M^2$ Paxos, ZooKeeper, Tango, Calvin, Spanner, and E-Paxos.
The differences in the system architecture and the underlying algorithms are also reflected in the methodologies for the system scalability experiments. $M^2$ Paxos used 100\% access locality to eliminate the overhead due to ownership migration and tested for up to 49 servers, while ZooKeeper uses up to 13 replicas for varying the write ratio from 0\% to 100\%. On the other hand, Tango evaluates for up to 100 servers since its scalability heavily relies on the underlying shared log, CORFU~\cite{Corfu}.  

\paragraph{Number of Regions}
\label{S:NumberOfRegions}
The number of regions of the deployment is a crucial factor for the wide area network systems due to the high network communication overhead. As mentioned in \ref{Setups}, in this survey, only the real wide-area network environments or testbeds with compatible network latencies are considered separate regions. Although they are deployed over wide area networks, systems with a fixed size of deployments like WPaxos, WanKeeper, and ZooNet are not included for scalability evaluations for the number of regions due to the lack of sole measurements for different numbers of regions. Omni-Paxos evaluates how it reacts to node failures and network partitions with nodes in a fixed number of regions, but does not discuss regional scalability. Among all the systems studied, only E-Paxos and Mencius purely analyze the system scalability for the number of regions. Mencius uses the fixed network latency by changing its regular topology from flat to star and having a central node to coordinate all network communications, while E-Paxos uses EC2 placement groups for the deployment with varying network latencies.

Most systems studying scalability concentrate on workload scalability or horizontal system scalability; however, there are some exceptions. For instance, $M^2$ Paxos also studies the vertical scalability of the system by varying the CPU performance at the nodes. Similarly, PNUTS evaluates the effects of disk space on the average latency. 
Some systems evaluate scalability differently due to architectural variations. For instance, DistributedLog focused on scalability with respect to the number of streams being processed, since a stream is a basic unit of sharding in the system.  However, because each physical node can only handle a limited number of streams, scaling the workload to include more streams ultimately causes the underlying scheduling system to span additional worker nodes to handle the streams. DistributedLog evaluation addresses this by performing workload scalability benchmarks concerning a single proxy to study the scalability limitations of a single node while doing system scalability on a cluster that can span additional workers.

\subsubsection{\textbf{Consistency}}
\label{S:Consistency}
\hfill\\ 
Evaluating the consistency properties of a protocol or an algorithm is no trivial task. Part of the reason is understanding the consistency itself, since many different consistency models exist. The ambiguity of consistency definitions also makes it difficult to compare the consistency guarantees provided by various systems without careful examination of the algorithms and protocols. For instance, many different systems claim strong consistency for their protocols; however, the actual guarantees provided may differ drastically, based on the transaction modes, assumptions about command ordering (total order, partial total order, etc), and assumptions about the client interaction with the system. 


Although benchmarking serializability or linearizability is not simple, testing for these guarantees is feasible. Bizur and Multi-Paxos perform such tests to detect any inconsistencies in their strictly serialized write operations. Bizur uses Serialla, a testing tool for strict serializability in the Elastifile file system \cite{Elastifile}, which produces concurrent updates while checking the responses at all replicas. It detects the requests causing inconsistent execution orders at any replica and provides a descriptive log of operations. Similarly, Multi-Paxos uses runtime checking for any inconsistencies at any replica by periodically sending checksum requests to all replicas. Replicas calculate the checksum value of their fault-tolerant log and compare that with the master's value. Jepsen \cite{jepsen} carries out a test on ZooKeeper to confirm the linearizability variant maintained under network partition and leader failure by partitioning the site that has the leader and another replica and keeps sending the write requests during partition. Then, it recovers the region and checks the logs at each replica. Elle \cite{Elle}, built over Jepsen, efficiently checks for violations of ScalarDB's strict serializability guarantees by constructing transaction dependency graphs and identifying critical dependency cycles.

In our study, we found very few systematic evaluations for consistency. However, some systems evaluate their consistency in terms of data {\bf staleness} at different nodes or geographical regions. Such staleness-based evaluations aim to show that clients cannot read stale or old values of data, no matter which node is being used for reading the data. In this manner, ZooNet measures the performance degradation while disabling stale reads by synchronizing read requests first with the owner of the requested data object to make sure it serves the most up-to-date data.

\subsubsection{\textbf{Availability}}
\label{S:Availability}
\begin{table*}[h]
\centering
\begin{tabular}{l|l|l|l|l|l|l|}
\cline{2-6}
& \multirow{2}{*}{\textbf{System}} 
& \multirow{2}{*}{\textbf{\# of Servers}}
& \multirow{2}{*}{\textbf{\# of Failures}}
& \textbf{\Longunderstack{Failed Node(s) \\Type}} & \textbf{\Longunderstack{\# of \\Regions}} \\
\hline \hline 
\multirow{7}{*}{\textbf{Alg.}}
& Mencius~\cite{Mencius} & 3 & 1 & Leader-Replica & 3  \\
&Raft~\cite{RAFT} & 5,9 & 1 & Leader & 1 \\
&E-Paxos~\cite{E-paxos} & 3 & 1 & Leader-Replica & 1 \\
&Bizur ~\cite{Bizur}  & 3 & 1 & Leader & 1\\
&WPaxos ~\cite{WPaxos} & 15 & 1 & Leader \& Follower &5 \\
&Omni-Paxos ~\cite{omnipaxos} & 5 & 1-2 & Leader \& Follower & 3 \\
&Hydra ~\cite{Hydra}&3 &1 &Leader &1 \\
\hline
\multirow{4}{*}{\textbf{C.S.}}
&ZooKeeper ~\cite{Zookeeper} &5 & 1-2 &Leader \& Follower &1\\
&FlexLog ~\cite{Flexlog} &3 &1 &Leader-Replica &1\\
&SplitFT ~\cite{SplitFT}&3 &1-2 &Leader-Replica &1 \\
&Narwhal ~\cite{Narwhal}&10 &1, 3 &Leader-Replica &5 \\
 \hline
\multirow{1}{*}{\textbf{App.}}
&Spanner ~\cite{Spanner} & 25 & 5 & Leader \& Follower & 5\\

\hline
\end{tabular}
\caption{Systems evaluating the availability and key parameters used (Alg: Algorithms; C.S: Coordination Services; App: Applications).}
\label{tab:availability}
\end{table*}

Availability evaluation usually involves benchmarking the system's ability to continue an operation in the presence of faults. The availability is often measured in terms of throughput degradation caused by the failure. Many kinds of failures are possible within the system or protocol, but the researchers in the distributed coordination community tend to concentrate on a {\bf crash-fault model} of operation and most often evaluate for node crashes.

Some other failure types, such as network partition around the node, may be indistinguishable from crashes for many protocols. Partitions, however, may cause different behavior than crashes in a few cases. For instance, a ZooKeeper ~\cite{Zookeeper} follower partitioned from the rest of the cluster can serve stale reads to the clients for some time, thus it will contribute to the throughput measurement. 

The ability to tolerate failures and remain available is one of the properties of consensus algorithms and coordination systems built on top of such algorithms. The number of failures a system can mask often depends on the cluster size, and larger clusters can mask more failures. In Table~\ref{tab:availability}, we summarize the vital parameters used by the systems when benchmarking for availability and fault tolerance. 
Most systems we have reviewed assumed no concurrent failures; however, ZooKeeper evaluation performed a benchmark with two follower nodes failing at almost the same time. The cluster size used for that evaluation of ZooKeeper allowed a maximum of two failures at the same time. 

Spanner availability evaluation is different from the rest of the systems and protocols. Spanner uses Paxos groups as a unit of data sharding and replication. In the availability evaluation, they created 1250 Paxos groups on 25 nodes in 5 regions, with each Paxos group taking a single machine from each region. Then an entire region has been crashed; however, since the Paxos protocol can mask up to 2 failures in a cluster of 5 nodes, all groups were able to continue operation. As a result, this benchmark is similar to the evaluation of availability on a single Paxos cluster of 5 nodes by crashing a node.  

On the other hand, Hybrid Paxos and Tango approach availability evaluation differently and evaluate the effect of increased availability with respect to performance. Hybrid Paxos increases the number of replicas under the same load and measures the average latency to see the impact of having more replication in the system. Similarly, Tango uses a primary-backup scenario for the same view of the Tango object, serves all read requests from the backup replica and all writes from the primary replica, and measures the throughput accordingly.


\section{Benchmarking Requirements for Distributed Coordination Systems}
\label{S:requirements}
Out of many reviewed coordination systems, protocols, and applications requiring distributed synchronization, only a handful of them (\cite{Tango, Calvin, Wankeeper, ryabinin2024swiftpaxos, cockroach, oceanbase, ScalarDB}) use a standard benchmarking suite or tool for evaluation. However, most of the works share a great deal of commonality in terms of what aspects of the system's behavior they evaluate. Performance evaluation is by far the most common type of benchmarking performed, while many of the authors also show fault tolerance or availability of their systems by measuring performance degradation caused by failures. 

The abundance of evaluations measuring the same aspects of system behavior and the lack of a prevalent benchmarking suite suggest that existing benchmarking tools are either deficient in covering some aspects of evaluated metrics or are not universally adaptable. Identifying and addressing the requirements for a complete evaluation of distributed coordination systems will allow us to customize existing tools accordingly or to create a new benchmarking suite that can be used by a wide range of applications exposing similar interfaces to the user. In the remainder of this section, we will discuss benchmarking challenges and the general requirements of a benchmarking suite for distributed coordination systems.

\subsection{Benchmarking Suite Flexibility and Sophistication}
Many systems resort to performing their evaluation with their custom benchmarking tools to showcase their strong features on particular workloads. This means that most systems require a high level of tenability from the benchmarking suite they are using. Consequently, if no popular suites can be adjusted to generate desired workloads, authors are forced to either modify the tools or make their own. 

A flexible benchmark should be highly customizable throughout the range of parameters, but it also needs to allow for benchmarking various metrics of the system's behavior. As such, a benchmarking suite for coordination systems needs to be able to evaluate all facets of the system's behavior: performance, scalability, availability, and consistency. Some of these metrics are related; for instance, the benchmarking suite can evaluate the system's scalability by measuring performance on a different system or workload. Some other metrics are orthogonal to each other, as is the case with performance and consistency.

{\bf Performance Benchmark.} For general performance measures, the benchmarking suite must provide the ability to manipulate various workload parameters: read-to-write ratio, size of the data pool, size of an individual object, data access overlap, access locality, and per-node workload distribution. 

{\it Read-to-write ratio} is a fundamental property of workloads for coordination systems and applications relying on distributed synchronization. The number of read and write operations a system is performing can vary greatly depending on the application. Many workloads are read-oriented; however, some tasks, such as logging, perform more updates.

{\it Size of data pool and size of individual object} control the overall size of the workload. Large object size can drastically increase the latency and reduce the throughput of the system; however, some systems optimize for larger objects while others work best with smaller data items. The size of the data pool is also used for generating variations for the percentage of conflicting commands, along with the data access overlap ratio. 

{\it Data access overlap} is important in the context of coordination services and coordinated applications. This parameter controls the likelihood of the same object being accessed by two or more different clients and hence it affects the command conflict ratio. In case of no overlap in data access, clients will never access objects belonging to other clients, while in a 100\% overlap, all clients are equally likely to access any data. This measure is especially important for multi-leader coordination systems or applications that allow concurrent writes, where conflicting commands require a special resolution and a longer time to complete. This is also closely related to consistency models for those systems, since for systems providing synchronous reads, all dependent requests need to be serialized. This is even more challenging for WAN deployments and has a significant impact on the latency evaluation. 

{\it Access locality} determines the distribution of the access patterns of clients. It is an important parameter, especially for systems using ownership mechanisms to satisfy their consistency guarantees. It has a significant effect on their performances as it either requires command forwarding in static ownership cases or ownership migration in systems using dynamic ownership.

{\it Per-node workload distribution} parameter allows controlling how much workload each system node receives. Controlling these distributions can allow the benchmark to create high-stress and low-stress regions in the system. It also provides the ability to evaluate the load-balancing capabilities of the system. Additionally, controlling which nodes process what commands is essential for enforcing the conflict rate.  

{\bf Scalability Benchmark.} In addition to workload parameters, the scalability benchmarking suite needs to have the ability to tune the amount of work it pushes through the system under test. This is often achieved by increasing the {\it number of concurrent clients} interacting with the systems and/or increasing per-client command throughput. The size of the cluster, the number of servers, and the geo-distribution level are not the parameters of the benchmarking tool. However, the benchmarking tool should provide information about throughput/latency per node as well as the cumulative throughput/mean latency, which enables comparing results from multiple runs of the benchmark with different configurations of the overall system architecture.  

{\bf Availability Benchmark.} Availability measurements require simulating system failures while measuring the performance of the system during them. Even though many types of failures are possible, all of the studied systems that underwent availability evaluation resorted to a crash-failure scenario. The ideal benchmarking suite needs to provide the ability to evaluate the system's behavior not only under crash-failures, but under other types of malfunctions, such as network partitions, clock drifts, memory corruption, and unreliable links. For many of these failures, the benchmark needs to control the {\it number of simultaneous failures}. 

{\bf Consistency Benchmark.} Consistency evaluations significantly differ from the prior three benchmarks. Most importantly, consistency is not evaluated through performance observations. Existing attempts at consistency evaluations often focus on studying data staleness. Staleness describes the amount of outdated data that can be read by the client. This is often good enough to show the eventual consistency guarantees of the system, but not nearly enough to test for all spectrums of possible consistency guarantees. This is especially important in the realm of consensus algorithms and coordination systems and applications using these algorithms, since such systems aim to provide stronger levels of consistency than eventual. In this matter, a consistency benchmark is required to evaluate the consistency guarantees provided by the system for different levels of linearizability and serializability under various combinations of client request ratios of read and write operations.

\subsection{Benchmarking Suite for WAN Systems}
Many of today's distributed applications are deployed on scales that span multiple datacenters across the country, region, or even the globe. Such a scale introduces many challenges that are not observed in a single-datacenter setting. Large distances between components of the system drastically increase communication delay and thus the system's latency. Cross-datacenter bandwidth may also be limited, driving performance degradation further. 

Many WAN systems (such as \cite{Wankeeper,E-paxos,WPaxos}) also see the performance and scalability artifacts from the geographical placement of data centers. On a global scale, it is no longer possible to assume roughly uniform latencies between nodes located in different datacenters. Physical distances start to dictate the speed of communication between the regions; thus, in a WAN system, the communication latency between regions can easily differ by the order of magnitude. These disproportional delays introduce penalties for some regions while giving benefits to others. In this matter, the distribution of the clients also becomes important since distributed clients may cause a variance in latency measurements. For instance, systems allowing their clients to communicate with any replica at any region would differ in latency from the systems limiting client communication to local replicas.  

These WAN challenges allow systems engineers to optimize for a wider range of workload parameters, such as {\it data locality} and {\it access locality}. Data locality controls the initial data distribution in the WAN system. Similarly, access locality is a measure of access pattern to the data objects shared and possibly replicated globally. 
A successful benchmarking suite must be able to generate workloads with these parameters in mind for WAN systems. 

\subsection{Benchmarking Suite Scalability}
The scale of modern systems is rather large. The protocols and simple coordination systems built on top of these protocols can easily span into tens of nodes, while the applications scale even further. The large scale of the application means that it can handle a lot of traffic, thus requiring a benchmarking tool that can scale with the system and put out an adequate workload.  

Benchmarks that do not scale will not be able to saturate larger systems, and will not provide a complete picture of those systems' performance and scalability. 
%
A typical way to scale a benchmark is to make it run multiple clients interacting with the system. However, often the benchmarking tools are limited to spanning the clients as separate threads ~\cite{YCSB}. Generating a workload out of a single machine may not always be enough to saturate large systems running in a cluster of many nodes. We believe that a benchmarking tool for distributed coordination systems needs to be distributed as well to scale well with the system under test.

Scaling a benchmarking tool to multiple machines is essential for proper WAN benchmarking. Since a suitable benchmark needs to control such parameters as the locality of the data and the locality of access, WAN systems must have at least one benchmarking node present in each region. However, making the benchmarking tools distributed over multiple nodes is not without challenges. For instance, the benchmarking nodes require some degree of synchronization to facilitate such tasks as starting and stopping the workloads, agreeing on the workload distributions across benchmarking nodes, and aggregating the results.

\subsection{Benchmarking Suite Ease of Adoption}
The ease of use and adaptability of a benchmark is a big contributing factor to many systems deciding not to adopt any of the standard benchmarks for their evaluations. Straightforward integration for various systems developed with different programming languages and frameworks is critical for any benchmarking suite. It is also vital that the benchmarking suite operates as a black box and does not require users to learn about the internals of the benchmarking suite. Similarly, the benchmarking suite should be configured to operate independently regardless of the evaluated system details, such as the programming languages used for the development. This is also an important factor for a fair comparison of the evaluated systems.

\section{Existing Benchmarks and Their Deficiencies}
\label{S:4}

In this section, we study several popular benchmarking frameworks for distributed systems, with the metrics they cover and the parameters they provide. Table \ref{tab:toolsmetric} summarizes the standard benchmarking tools used in the studied systems and some other state-of-the-art benchmarks that could be utilized for distributed coordination systems evaluation. As shown in the table, none of the standard benchmarks can accommodate all aspects of the comprehensive evaluation of distributed coordination systems. This explains why researchers tend to create their ad-hoc benchmarks, as can be seen in table \ref{Setups}. Another approach could be combining these standard benchmarks to cover all aspects, such as using YCSB for performance and Jepsen for availability and consistency evaluations. However, YCSB comes short for distributed environments and does not support some important parameters for the evaluations, such as data access overlap and access locality. Similarly, Jepsen does not provide black box benchmarking and requires expertise in Jepsen tools. Elle, implemented over Jepsen, provides substantially more efficient isolation checks for black box databases. We will analyze some of the state-of-the-art benchmarking tools in detail under their main evaluation category.

\begin{table*}[]
\centering
\resizebox{\textwidth}{!}{%
\begin{tabular}{|c|c|c|c|c|c|c|c|}
\hline
\multirow{2}{*}{\textbf{Tools}} & \multicolumn{2}{c|}{\textbf{Performance}} & \multirow{2}{*}{\textbf{Scalability}} & \multicolumn{2}{c|}{\textbf{Availability}} & \multicolumn{2}{c|}{\textbf{Consistency}} \\ \cline{2-3} \cline{5-8} 
 & \textbf{Throughput} & \textbf{Latency} &  & \textbf{Node Failure} & \textbf{Network Partition} & \textbf{Staleness} & \textbf{Linearizability} \\ \hline \hline
YCSB\cite{YCSB} & \cmark & \cmark & Single Client &  &  &  &  \\ \hline
YCSB+T\cite{ycsb+t} & \cmark & \cmark & Single Client &  &  &  &  \\ \hline
YCSB++\cite{ycsb++} & \cmark & \cmark & Distributed &  &  & \cmark &  \\ \hline
TPC-C\cite{TPC-C} & \cmark & \cmark & Distributed &  &  & &  \\ \hline
BG\cite{bg} & \cmark & \cmark & \cmark &  &  &  &  \\ \hline
UPB\cite{upb} & \cmark & \cmark & Distributed & \cmark &  &  &  \\ \hline
Chaos Monkey\cite{chaosmonkey} &  &  &  & \cmark & &  &  \\ \hline
HiBench\cite{hibench} & \cmark & \cmark & \cmark &  &  &  &  \\ \hline
BigDataBench\cite{BigDataBench} & \cmark & \cmark & \cmark &  &  &  &  \\ \hline
Jepsen\cite{jepsen} &  &  & Single Client & \cmark & \cmark & \cmark & \cmark \\ \hline
Elle\cite{Elle} &  &  & Distributed & & & \cmark & \cmark \\ \hline
BenchFoundry\cite{benchFoundry} &\cmark  &\cmark  &\cmark (Distributed) & & & \cmark & \\ \hline
\end{tabular}%
}
\caption{Metrics provided by benchmarking tools.}
\label{tab:toolsmetric}
\end{table*}

\begin{table*}[]
\centering
\resizebox{\textwidth}{!}{%
\begin{tabular}{|c|c|c|c|c|c|c|}
\hline 
\textbf{Tools} & \textbf{\# of Objects} & \multicolumn{1}{l|}{\textbf{Size of Objects}} & \textbf{Read/Write Ratio} & \textbf{Data Access Overlap} &\textbf{Access Locality} & \textbf{\# of Clients} \\ \hline \hline
YCSB\cite{YCSB} & \cmark & \cmark & \cmark &  & & Single Process \\ \hline
YCSB+T\cite{ycsb+t} & \cmark & \cmark & \cmark &  & & Single Process \\ \hline
YCSB++\cite{ycsb++} & \cmark & \cmark & \cmark &  & & Distributed \\ \hline
BG\cite{bg} & \cmark & \cmark & \cmark & \cmark & & Distributed \\ \hline
UPB\cite{upb} & \cmark & \cmark & \cmark &  & & Single Process \\ \hline
Jepsen\cite{jepsen} & \cmark &  & \cmark &  & & Single Process \\ \hline
BenchFoundry\cite{benchFoundry} & \cmark &  & \cmark &\cmark  &\cmark & Distributed \\ \hline
\end{tabular}%
}
\caption{Parameters tunable by benchmarking tools.}
\label{tab:toolspara}
\end{table*}

\subsection{Performance}
Performance evaluation is the most supported aspect of benchmarking tools for distributed coordination systems, as can be seen in Table \ref{tab:toolsmetric}. Performance evaluations are prone to be sensitive to workload characteristics. To observe the true performance of a distributed coordination system, benchmarking tools allow the configuration of the parameters listed in Table \ref{tab:toolspara}.

\textbf{YCSB} \cite{YCSB} has become a standard benchmarking tool for the evaluation of NoSQL frameworks since its publication in 2010. The main reason for its popularity is generality and extendability. YCSB is a general-purpose benchmarking tool and can be used for all NoSQL database systems. It is easy to extend to any additional database by implementing a simple CRUD (create, read, update, and delete) plus scan interface against the datastore under benchmarking. Although the YCSB workload does not model real applications, its generated synthetic workload is highly tunable in five dimensions: the number of clients, operation ratio, request distribution, key space, and throttled throughput. Even though it enables the distribution of the requests over the key space to some level, it does not provide the requirements for evaluating access locality since it supports only a single-client implementation. Additionally, despite the workload model supporting uniform, ziphian, latest, and multinomial distributions, it comes short of precise adjustments for data access patterns, and it requires customization to set up with certain probabilities. Likewise, YCSB ignores creating conflicting commands. These parameters limit even the performance evaluations since systems have different consistency models, and they have a significant impact on latency. For instance, a coordination system using a data ownership mechanism may need access migration for some data objects. It is simply not provided by any single-client benchmarking solution. 
YCSB only measures the performance metric in our category. In particular, it measures the latency of each operation and the overall throughput. Another limitation of YCSB is a lack of scalability for the WAN setting since YCSB only generates workload from one process. Our experience from benchmarking a geo-distributed system makes us realize that YCSB does not span multiple datacenters and cannot generate workloads with some locality characteristics.


\textbf{YCSB+T} \cite{ycsb+t} is an extension of YCSB that adds a new tier of transactional operations and validation. To generate a meaningful workload, YCSB+T defines the Closed Economy Workload, which simulates bank account transactions with a fixed total amount. It supports transactional read, scan, update, delete, and readModifyWrite operations. The readModifyWrite operation includes reading two data objects and updating both. The validation phase of the benchmark tries to detect anomalies by comparing the total account balance before and after transactions. Given the database state, this validation cannot detect any dirty reads or lost updates that do not change the sum of all account balances. YCSB+T mimics the workload features of YCSB, and it suffers from similar limitations. It is not well-suited for WAN deployments and evaluating WAN systems, as it does not support the distribution of the clients, hence producing conflicting commands and managing access locality. 

Similarly, \textbf{YCSB++} \cite{ycsb++} extends YCSB to support multiple clients that can run on different machines. To manage synchronization and group membership of clients, it uses ZooKeeper as the coordination service and the notification mechanism. It occupies HBase \cite{HBase} and Accumulo \cite{Accumulo} as its table stores. To provide more realistic evaluations, it enables some other features of table stores, such as table pre-splitting for fast ingest, server-side filtering, and bulk loading of the data. As another aspect of table stores, it enables evaluating the effect of applying access control on performance by using pre-configured access control lists (ACLs) to map credentials and operations for any schema. By using a monitoring tool, Otus \cite{otus}, performance metrics of table stores and YCSB++ clients are collected in a central repository for fine-grained analysis. However, YCSB++ is tailored for table stores and is more suited for evaluating big data applications. Moreover, the synchronization of the clients relies on the ZooKeeper service, which performs poorly on the WAN scale. Although it supports the distribution of the clients, the distribution of the workload does not provide the ability to adjust data access overlap or access locality parameters. 

\textbf{BG} \cite{bg} is another benchmark that mainly focuses on performance under a specific real-world application workload. It models the workloads of social networking applications that have read operations like listing all friends or reading top posts for a user, and write operations such as accepting friendship invitations. BG summarizes the performance evaluation in terms of Social Action Rating based on the customizable service level agreement (SLA). SLA is defined by four parameters, namely the percentage of requests to observe less than pre-specified response time, the response time, the unpredictable data amount, and the unpredictable time limit for any data object. This simply models the measurements for latency and staleness. BG workload originally creates overlapping data access patterns since clients may request conflicting actions, such as one client requesting acceptance of friendship invitations and another client requesting the rejection of the same invitation. However, BG uses a locking mechanism to avoid these conflicting commands. Similarly, although the clients can be distributed in BG, it only uses this for scalability measures and does not accommodate the need for access locality parameters.  

\textbf{BenchFoundry} \cite{benchFoundry} adopts a different approach to create workloads. Instead of providing predefined workloads, it allows the creation of the desired sequence of operations. In this way, it could be possible to generate custom workloads for the parameters listed in Table \ref{tab:toolspara}. However, it uses master-slave architecture before running the benchmarking clients, which determines the order of the execution of the requests listed in each trace file. If we assume the jitter in clock synchronization is minimized, this would eliminate the conflicting commands even if clients can be configured to share the same key space. 

\textbf{HiBench} \cite{hibench} aims to provide a more comprehensive and realistic evaluation of Hadoop with various pre-defined workloads such as web searches, machine learning tasks, and HDFS operations. The workloads offer different object sizes and data access patterns, but are not effective in evaluating conflicting operations by nature of the Hadoop workflow, and apply only to Hadoop-like systems. \textbf{BigDataBench} \cite{BigDataBench} expands on this with more workloads and targeted systems, but does not allow users to tune conflicts.

\subsection{Availability}

\textbf{UPB} (Under Pressure Benchmark) \cite{upb} is the benchmark for measuring and quantifying the availability of distributed database systems. UPB uses distributed YCSB workloads and a more complex evaluation scenario with different load sizes, replication factors, and the number of failed nodes. UPB aims to measure the performance impact on the system before, during, and after node failures.
UPB leverages the YCSB workload generator for its availability evaluation and therefore inherits all its limitations too. That being said, UPB can be a guide for any benchmarking suite on quantifying and comparing availability.

\textbf{Chaos Monkey} \cite{chaosmonkey} is a tool that randomly terminates virtual machine instances and containers that run inside a cloud. Such random node termination could potentially reveal any problem in distributed systems, thus ensuring that engineers implement their services to be resilient to instance failures.
Chaos Monkey is a good example that follows chaos engineering for fault injection in any benchmarking suite. Chaos Monkey lets the user define their own outage checker for availability checking instead of giving any performance or other evaluation as a plugin.

\subsection{Consistency}

The benchmarking frameworks that try to evaluate consistency, like \textbf{YCSB++} \cite{ycsb++}, can usually measure only the read operation staleness in a weakly consistent system. YCSB++ coordinates multiple distributed clients using ZooKeeper and measures time-based staleness. The write operation is published to ZooKeeper right after it is completed. Then any subscribing client can get the written value and repeat read operations until it reads a new value. The time difference between the first and last read is the approximated lower bound of staleness. One benefit of such a method is the ability to measure consistency online, and the longer the benchmark runs, the better the chance to minimize the coordination noise. Similarly, BG measures the unpredictable data amount as it constrains the synchronization of the data with an unpredictable time amount in the SLA. To measure this, BG logs each read and write request with unique IDs, and then it compares their values based on the timestamp of the operation. This approach brings the accuracy for the time synchronization in the distributed setting of benchmarking clients. BenchFoundry follows a similar approach by logging all fine-grained results to evaluate staleness. On the other hand, it relies on the clock synchronization of the clients, which is challenging in distributed settings. Out of the listed benchmarking tools in table \ref{tab:metrics}, only \textbf{Jepsen} \cite{jepsen} and \textbf{Elle} \cite{Elle} provide testing for consistency models in terms of linearizability and serializability. Jepsen does not provide a black box benchmarking interface, and it requires customizing workload characteristics depending on the consistency models. It simply enables the user to customize the workload to check whether the proposed consistency guarantees are preserved. Elle, however, can evaluate isolation guarantees of a black box distributed database more efficiently than Jepsen's KNOSSOS \cite{Knossos} by constructing a dependency graph from transaction history and identifying violating cycles.

\section{Related Work}
\label{S:relatedWork}
Distributed systems have been maintaining their importance for the last several decades due to the increase in the need for scalable and reliable distributed applications while preserving high performance. 
To analyze distributed systems comprehensively and compare them in terms of features and services, various surveys and evaluations have been published in the past. Surveys on cloud providers, data warehouses, distributed file systems, or metadata services can be counted among them. 

Cloud providers are analyzed and evaluated in terms of elasticity \cite{CMART}, computing power \cite{comperative-benchmarking}, and cost to performance efficiency \cite{fair-benchmarking} in previous efforts. Widely used distributed services are also analyzed in many works, such as a survey on stream processing \cite{stream-benchmarking} or performance and dependability evaluation of MapReduce systems \cite{MapReduce-benchmarking}. Similarly, different aspects of distributed systems are studied in several surveys, like reliability analysis on distributed systems \cite{reliability-survey} and load balancing characteristics of known systems \cite{load-balancing-survey}.

As a big part of distributed systems, data warehouses and file systems are studied for many specifications. Evaluation of distributed data warehouses for the cost-effectiveness of different hardware configurations \cite{ALOJA} and query performance of distinct design choices\cite{benchmarking-data-warehouse} are among the known efforts in these works. Distributed file systems are examined in many past works for general concepts \cite{file-systems-concepts,file-systems-gen1} or specific applications such as distributed access control \cite{access-control-file-systems}. Due to the differences in optimization, design techniques, and the complex interactions between the file systems and other system components like the kernel or operating system, benchmarking distributed file systems is not trivial. To identify the important metrics for the evaluation of distributed file systems, researchers also studied benchmarking file systems \cite{File-system-benchmarking,benchmarking-file-rocket}. 

Analysis of distributed coordination services in terms of general characteristics and importance of coordination \cite{importance-of-coordination} and the comparison of existing algorithms \cite{paxos-made-simple} are among the published works. However, to the best of our knowledge, there is no published work on the evaluation of distributed coordination systems. As mentioned in the Introduction, due to the lack of standard benchmarking tools for distributed coordination services, developers widely use their ad-hoc benchmarks, which are prone to unfair comparisons or limited results for the evaluation of the systems. This study is unique in identifying the metrics and parameters for the evaluation of distributed coordination systems, discussing how each system uses these metrics and parameters for its evaluation, pinpointing the deficiencies of well-known benchmarking suites in evaluating distributed computing systems, and finally discussing the features of an ideal distributed coordination benchmark. 

\section{Conclusion}
\label{S:conclusion}
Increasing demand in web-based big-data applications brought out the need for efficient use of distributed systems, which highly depend on the adequate implementation of distributed coordination. The distributed systems community has developed different protocols, coordination services, and distributed applications built on top of these services to satisfy this rapid growth in big data applications. However, due to the lack of a standard benchmarking tool, developers generally opt to use ad-hoc evaluation mechanisms and microbenchmarks. Hence, the evaluation of these systems has been very limited, resulting in inadequate and misleading measurements and an unfair comparison of the competing systems. In this paper, we have analyzed and compared well-known and widely used distributed coordination services, their evaluation mechanisms, and the tools used to benchmark those systems. We have identified the essential requirements of distributed coordination service benchmarking, such as the metrics and parameters for the evaluation of the performance, scalability, availability, and consistency of these systems.

Despite the commonalities in tuning parameters like read/write ratio, data access overlap, or size of data objects for performance measurements, access locality and number of data objects are not extensively evaluated by most of the systems. These two parameters, on the other hand, may have a significant impact on the system performance for systems that allow concurrent writes, since they are the main factors in producing conflicting or colliding commands. Similarly, they are critical for managing ownership of data objects for systems using ownership mechanisms to resolve conflicts. While most of the systems are evaluated with a single value of read/write ratio and size of data objects, tuning these parameters for different values is essential to understand the system performance extensively for different types of workloads. Depending on the type of applications using these services, the percentage of update operations or the size of data objects may change dramatically. Likewise, data access overlap is evaluated only for systems sharing the entire data space, but it is also a system-specific parameter that needs to be analyzed for different scenarios. 


The benchmarks discussed in this work do consider some of these tuning parameters, but often do not prioritize them when evaluating coordination algorithms and services. Benchmarks on performance often neglect scalability, data access overlap, and request conflicts as seen in YCSB++ \cite{ycsb++}, BG \cite{bg}, and BenchFoundry \cite{benchFoundry}. Some benchmarks on availability neglect to test for node failures, clock drift, and network partition altogether, as seen with UPB \cite{upb} and Chaos Monkey \cite{chaosmonkey}. Linearizability is often challenging to evaluate and needs more attention. To overcome these limitations with the existing benchmarks, some systems (such as ScalarDB \cite{ScalarDB} and CockroachDB \cite{cockroach})
combines multiple existing benchmarks to achieve the desired functionality. Most of the other systems considered in this study prefer using their own specially tailored microbenchmarks instead.

\label{S:References}
\bibliographystyle{abbrv}
\bibliography{sample.bib}

\begin{IEEEbiography}
[{\includegraphics[width=1in,height=1.25in]{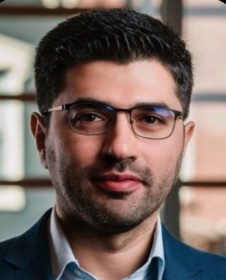}}]{Bekir Turkkan} is a research scientist at IBM Reearch. He received his Ph.D. in computer science and engineering from the University at Buffalo in 2024. His main research interests include distributed computing, scalability and performance of distributed systems and AI workloads, and energy-efficient system optimization.
\end{IEEEbiography}

\begin{IEEEbiography}[{\includegraphics[width=1in,height=1.25in]{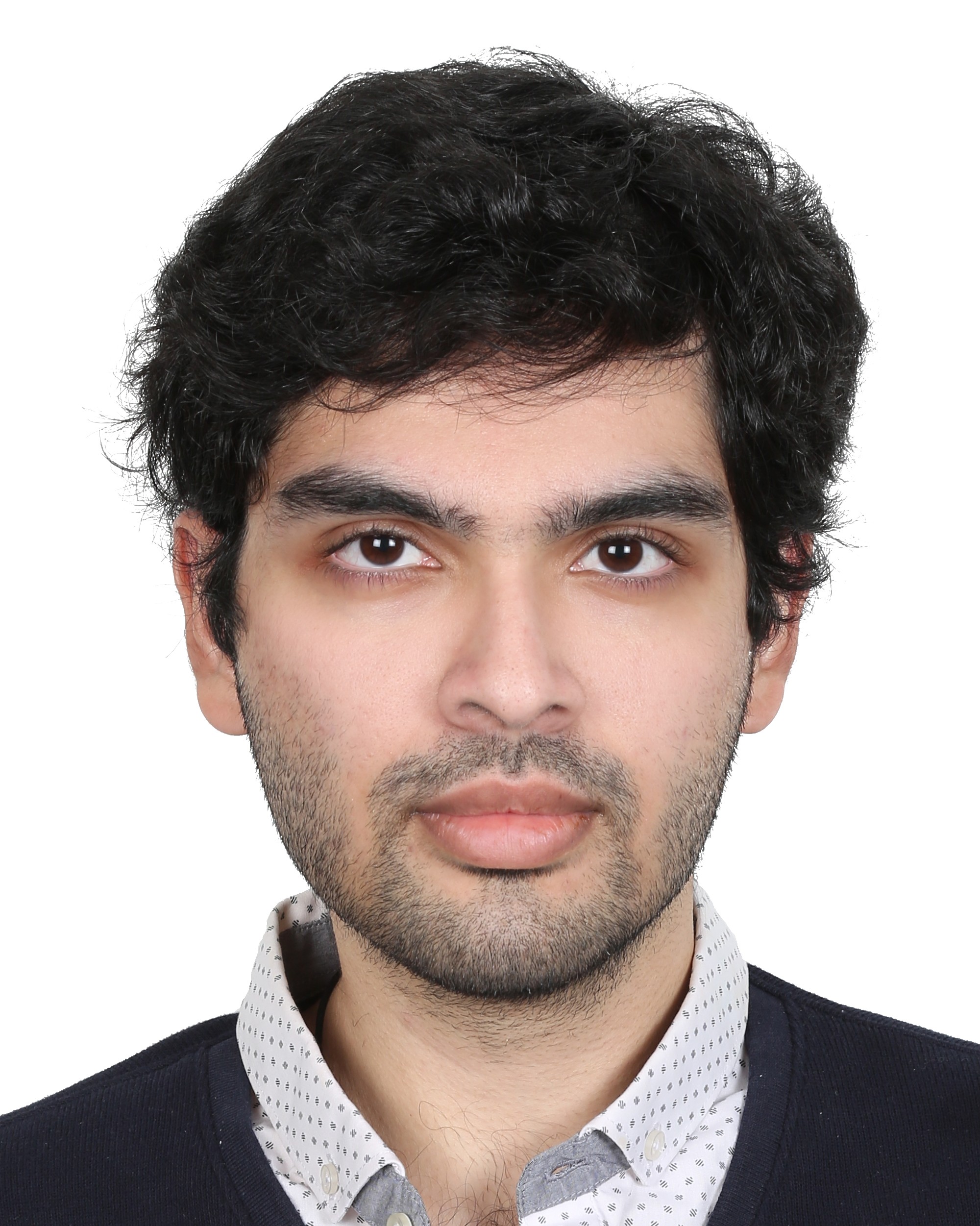}}]{Elvis Rodrigues} is a Ph.D. student in the Computer Science and Engineering Department at the University at Buffalo. His research interests include distributed systems, sustainable computing, scalability and performance of wide-area networked systems, and energy-efficient system optimization.
\end{IEEEbiography}

\begin{IEEEbiography}[{\includegraphics[width=1in,height=1.25in]{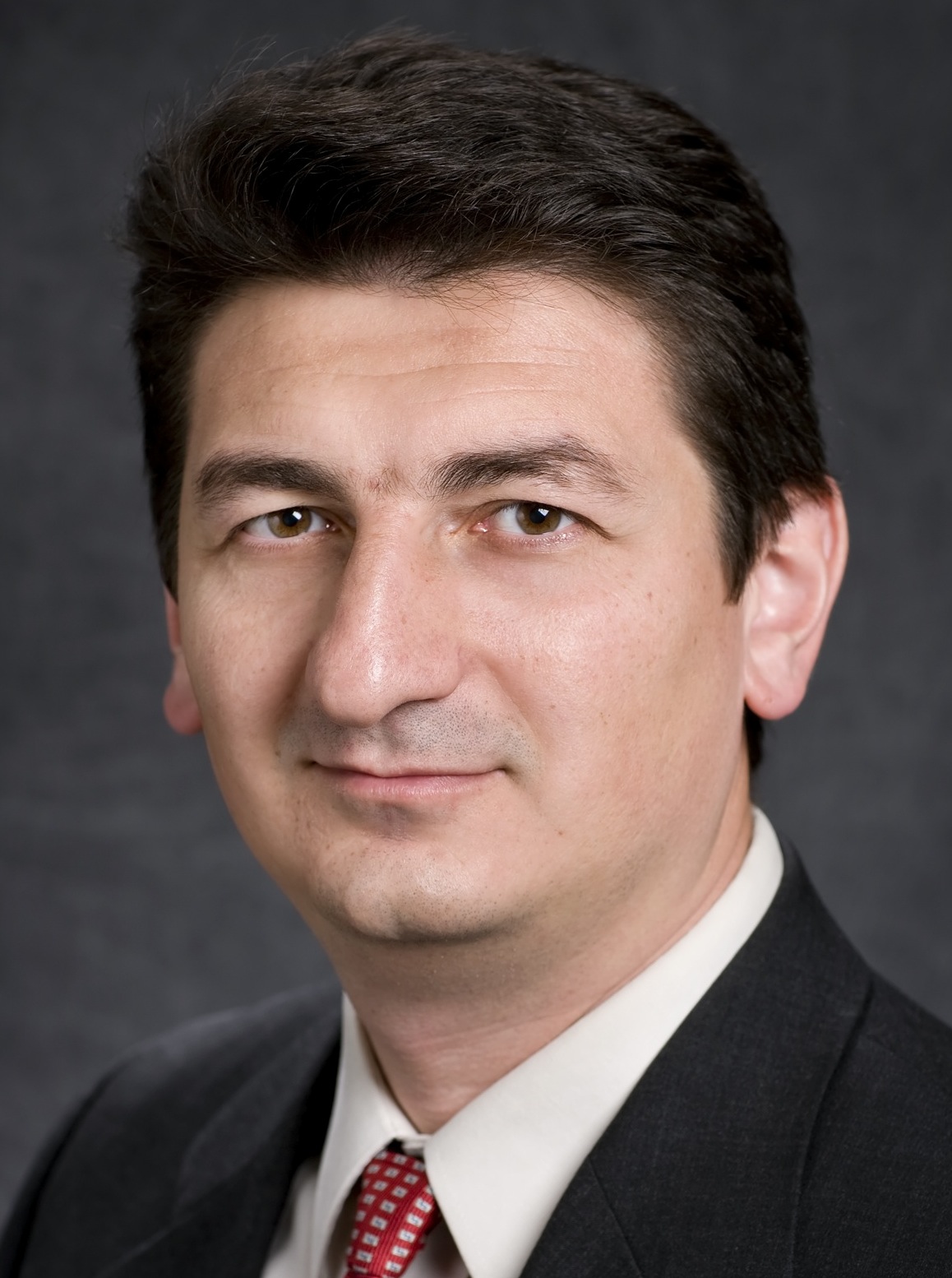}}]{Tevfik Kosar}
is a Professor in the Department of Computer Science and Engineering at the State University of New York at Buffalo. He received his Ph.D. in Computer Science from the University of Wisconsin-Madison in 2005. His main research interests include optimization of the performance, scalability, and sustainability of distributed systems, high-performance computing, and big-data analytics pipelines. Some of the awards received by Dr. Kosar include the NSF CAREER Award, IBM Research Award, Google Research Award, IEEE Region-I Technological Innovation Award, and UB Exceptional Scholar: Sustained Achievement Award. 
\end{IEEEbiography}

\begin{IEEEbiography}[{\includegraphics[width=1in,height=1.25in]{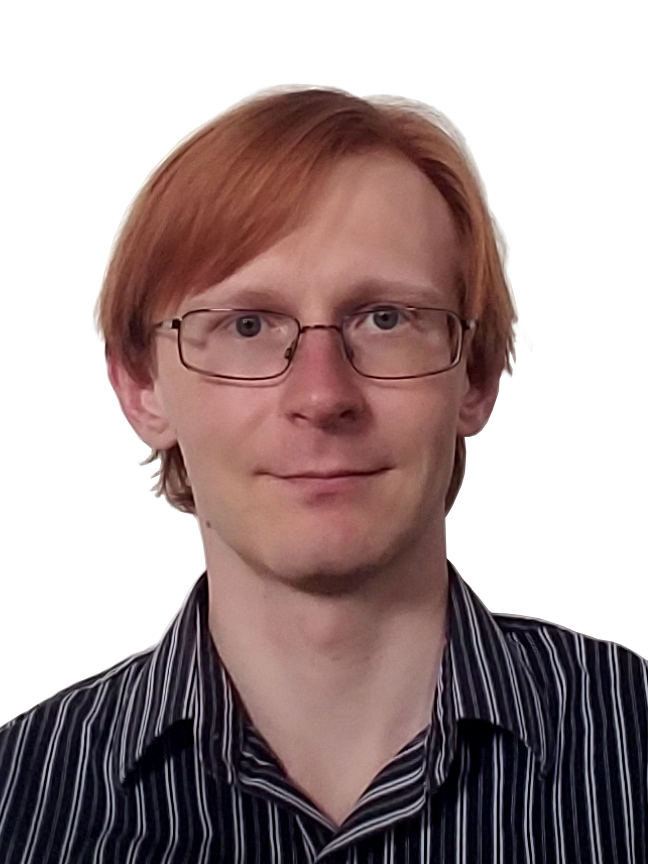}}]{Aleksey Charapko} is an assistant professor of computer science at the University of New Hampshire. He received his PhD in computer sciences from the University at Buffalo, and his B.S. and M.S. degrees from the University of North Florida. Dr. Charapko's main research interests are in the area of distributed systems, distributed consensus, distributed databases, and fault tolerance. Dr. Charapko is a Member of the IEEE.
\end{IEEEbiography}

\begin{IEEEbiography}
[{\includegraphics[width=1in,height=1.25in]{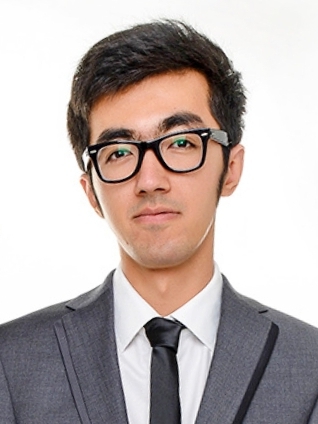}}]{Ailidani Ailijiang} is a researcher at Microsoft. He received his Ph.D. in computer science and engineering from the University at Buffalo in 2018. He received his B.E. in network engineering from Dalian University of Technology in 2006. His main research interests include distributed computing, fault tolerance, and consensus in networked systems.
\end{IEEEbiography}

\begin{IEEEbiography}[{\includegraphics[width=1in,height=1.25in]{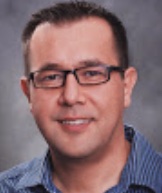}}]{Murat Demirbas}
is a Principal Scientist at MongoDB. He received his Ph.D. from Ohio State University in 2004 and did a postdoc at MIT in 2005. His research interests are in distributed and networked systems and cloud computing. Dr. Demirbas received an NSF CAREER award in 2008, the UB Exceptional Scholars Young Investigator Award in 2010, UB School of Engineering and Applied Sciences Senior Researcher of the Year Award in 2016. He maintains a popular blog on distributed systems.
\end{IEEEbiography}

\end{document}